# Discovery of Diffuse X–Ray Emission in 47 Tuc

Martin Krockenberger and Jonathan E. Grindlay

Harvard–Smithsonian Center for Astrophysics

## Abstract

We present the results of a search for diffuse x–ray sources in a 65 ksec ROSAT PSPC exposure of 47 Tuc. There is faint, soft emission on the NE side of the cluster at a distance of $6'$ from the core. The location of this emission along the direction of proper motion of the cluster suggests that it might be due to a bow shock. We show that a simple shock model fits the observed luminosity and temperature. However, a bow shock can only form if the rotation of the halo gas at the height of 47 Tuc (z = 3.3 kpc) is small compared to the rotation of the galactic plane. Therefore this observation provides not only the first x–ray detection of hot gas in a globular cluster, but also constrains the dynamics of the halo gas. We also find two sources of diffuse x–ray emission to the NE of the cluster which are brighter and harder. We consider a variety of models for this emission, including thermal emission from a high velocity wind from a hot white dwarf, and propose that it is due to inverse Compton emission from acceleration in the bow shock of low energy cosmic ray electrons from the population of millisecond pulsars in the cluster.

**Subject headings:** globular clusters – 47 Tuc; x–ray emission;

## 1 Introduction

The theory of stellar evolution of globular cluster (GC) stars (Renzini 1979), as well as comparison of the observed mean mass of white dwarfs (0.55 $M_\odot$) to the mass of low mass stars when they turn off the main sequence (0.85 $M_\odot$), indicate that Population II giants each lose about 0.2 – 0.3 $M_\odot$ (Roberts 1988). GCs, which contain large numbers of Population II giants, are expected to accumulate a few hundred $M_\odot$ of gas lost from giants between plane passages (Tayler and Wood 1975). Searches for H I (Bowers et al. 1979), H II (Hesser and Shawl 1977, Grindlay and Liller 1977) and dust (Gillett et al. 1988) have been conducted. The upper limits found in each of these searches are typically lower by a factor of 100 than what is expected. This suggests that there is a mechanism that constantly removes the gas from globular clusters. Several mechanisms have been suggested such as sweeping of globular clusters by halo gas (Frank and Gisler 1976), ∼100 km/sec winds from giants (Faulkner and Freemann 1977, Dupree et al. 1992) and UV driven winds (VandenBerg 1978). Recently Spergel (1991) has suggested that oblation from millisecond pulsars in GCs may expel some of the gas.

47 Tuc is an excellent object to test the different models of gas removal. It is one of the most massive clusters in the Galaxy ($1.3 \times 10^6 M_\odot$ (Pryor and Meylan 1993)) with a deep potential, so gas lost from giants is more likely to be retained in the cluster. Furthermore it is relatively nearby (4.6 kpc, Peterson 1993) which increases the chance of seeing emission from gas or gas interaction. The potential of 47 Tuc is deep enough that sweeping by halo gas is only effective in the outer parts (Frank and Gisler 1976). There will be some gas left around the core which could form a bow shock with the halo gas. Hartwick, Cowley and Grindlay (1982) analyzed deep pointings made with the Einstein Observatory and reported evidence for diffuse x–ray emission in 47 Tuc as well as in M22 and $\omega$–Cen. Confirming the presence of a bow shock in these clusters would prove that sweeping by the halo gas is effective in removing gas from GCs. Most of the other, less massive clusters would be swept free of gas by the halo. However, Faulkner and Smith (1991) showed that the diffuse x–ray emission reported for 47 Tuc is probably not in the direction of the relevant proper motion and that the x–ray luminosity is too large.

Evidence for the interaction between the ISM and gas in globular clusters has been found in the planetary nebula (PN) in M22 (Borkowski et al. 1993). This PN is stripped by the motion of the cluster through the halo. The difference between M22 and most of the other clusters is that M22 is very close to the galactic plane (z = 400 pc) and therefore encounters a much higher gas density ($n \sim 0.1 cm^{-3}$, Borkowski et al. 1993) than for example 47 Tuc, which is at z = 3.2 kpc where the halo gas density is estimated to be $n \sim 0.001 cm^{-3}$ (Savage and de Boer 1981).

In this paper we present the results of a new search for bow shock emission in a deep ROSAT PSPC image of 47 Tuc, which was obtained to follow up on the results of Hartwick, Cowley and Grindlay (1982). We find emission consistent with a bow shock and compare the results to theory. We also find other more prominent regions of diffuse, harder x–ray emission which we suggest may be inverse Compton emission from particles accelerated in the bow shock.

## 2 Data Analysis

We obtained and analyzed a deep (65 ksec) ROSAT PSPC image of 47 Tuc which was obtained during the first series (AO1) of pointed ROSAT observations (1991–1992) to find and understand diffuse x–ray emission from the cluster.

Figure 1 shows the bright central source complex at RA, dec (J2000) = ($0^h 24^m$, $-72°04.5'$) and the SE source ($0^h 25^m$, $-72°13'$) that was detected by EINSTEIN before. There is diffuse emission to the NE of the cluster center ($0^h 24.5^m$, $-72°0'$) which is radially aligned (see further discussion below) with the cluster center. The bright diffuse source at the southern edge of the image ($0^h 23.5^m$, $-72°24'$) is probably a background galaxy cluster (see below).

### 2.1 Data Reduction

To search for a bow shock, which we expect to be soft and faint (Faulkner and Smith 1991), we had to remove the background very carefully. The data were taken during 80 separate observation intervals of about 800 sec each in two (primary) periods (March, April 91 and April 92). We took the data from each of these intervals, divided it into ten equal time bins and measured the background rate in each of these bins. This was done separately in the soft (0.1 – 0.4 keV) and the hard (0.4 – 2.4 keV) bands. In the resulting background versus time plot we saw the gain change in the PSPC in October 1991. The change in gain is small compared to the frequent peaks in the background due to orbital background variations. We therefore determined the average background rate in our data and rejected all intervals above 2 $\sigma$. Of the total 61155 seconds, we rejected 8823 sec in the soft band and 8621 sec in the hard band.

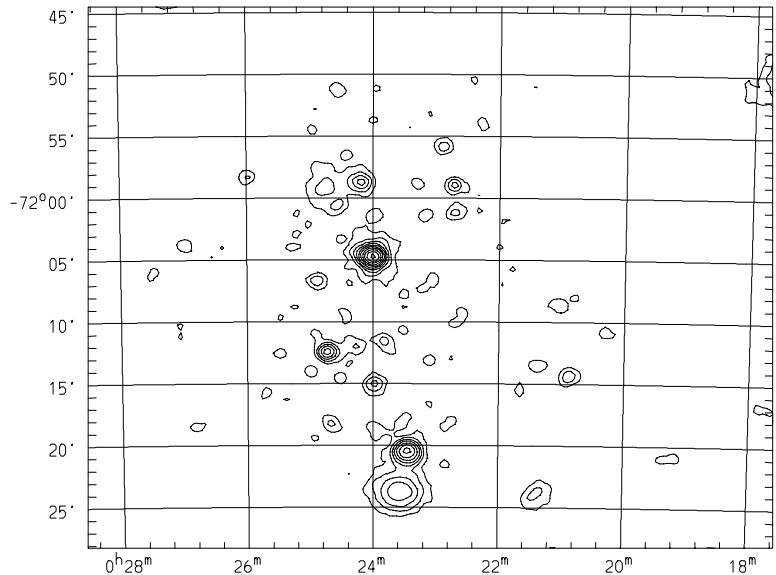

**Figure 1:** Raw ROSAT PSPC image of field centered on 47 Tuc in total energy band (0.1 – 2.4 keV). The cluster core source complex is at RA, dec (J2000) = ($0^h24^m$, $-72°04.5'$). The image is smoothed with a Gaussian of $\sigma = 15''$ and the contour levels start at $5.6 \times 10^{-3} cnts/pix$, which is about twice the value of the background. The contours increase logarithmically to 0.5 cnts/pix.

The second step was to remove all bright point sources from the image. This was necessary to be able to smooth the image on a scale of a few arcmin. We removed the 48 sources found by the automatic ROSAT PSPC processing (for the entire field) with the point spread function (PSF) given in Hasinger et. al (1992). The faintest sources found by the automatic processing each have a total of $\sim 20$ counts. As we are mostly interested in sources within $20'$ of the center of the image, we assumed that the PSF is not changing with position. We tested the quality of the removal of each point source by comparing the azimuthally averaged number of counts of the source to the PSF model. A K–S test showed that the PSF is a good fit to all the point sources, except for the central source, which is a complex of multiple sources.

We removed the central source complex with the four brightest core sources detected with the ROSAT HRI in 1992 as reported by Hasinger et al. (1994). Because the components of the central source are variable we changed the brightness of each of the four sources until all of the flux in the core was removed. We checked the removal by requiring the number of counts in each pixel after smoothing with a Gaussian ($\sigma = 20''$) to be less than the background plus its Poisson noise. This procedure removed the flux in the core (central $2'$), so we cannot measure the amount of diffuse emission in the core itself. By removing all the flux in the core we can be certain that smoothing does not spread counts from the core to regions further out.

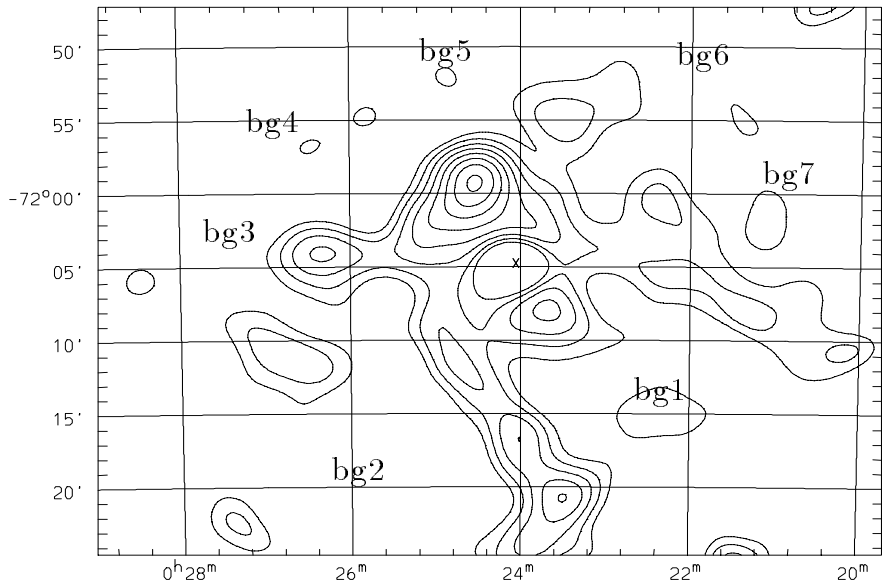

**Figure 2:** Final contour map of the soft band (0.1 – 0.4 keV) after point source subtraction and smoothing with $\sigma = 80''$. The contour levels start at $1.5 \times 10^{-4} cnts/pix$ and increase linearly by $8.3 \times 10^{-5} cnts/pix$ per contour to $7.3 \times 10^{-4} cnts/pix$, which is only $\sim 30\,\%$ of the mean sky background. The center of the cluster is marked with an "X". The central source complex has been removed to the level of the background. The seven positions where the background was determined are marked.

After removing all bright point sources, we smoothed the data on a large scale ($\sigma = 100''$) to examine the flatness of the background. We measured the background in the soft image in seven non–overlapping $4'$ by $4'$ boxes. These boxes are all at a distance of $10' - 15'$ from the center of the cluster, as shown in Figure 2. The fluctuations between the different boxes are less than 3 %. The value for the soft background is $2.7 \times 10^{-3} cnts/pix$ where one pixel covers $0.25\ arcsec^2$.

It was more difficult to do the same measurement in the hard image, because there still was a lot of structure after we removed the 48 point sources. Most of the structure is due to fainter point sources. Therefore we measured the background in only 2 positions, which were on the E and W side at $15'$ from the core. The two measurements differed by only 0.3 % . The value of the background we adopt in the hard band is $9.2 \times 10^{-4} cnts/pix$. Given these measurements we conclude that there are no significant large scale background fluctuations. None of the background measurements, however, cover the central $10'$ of the cluster, because of the diffuse emission there and the uncertain removal of the central sources.

## 2.2 Detected Diffuse Sources

### 2.2.1 Sources A and B: Bright, Compact Diffuse Sources

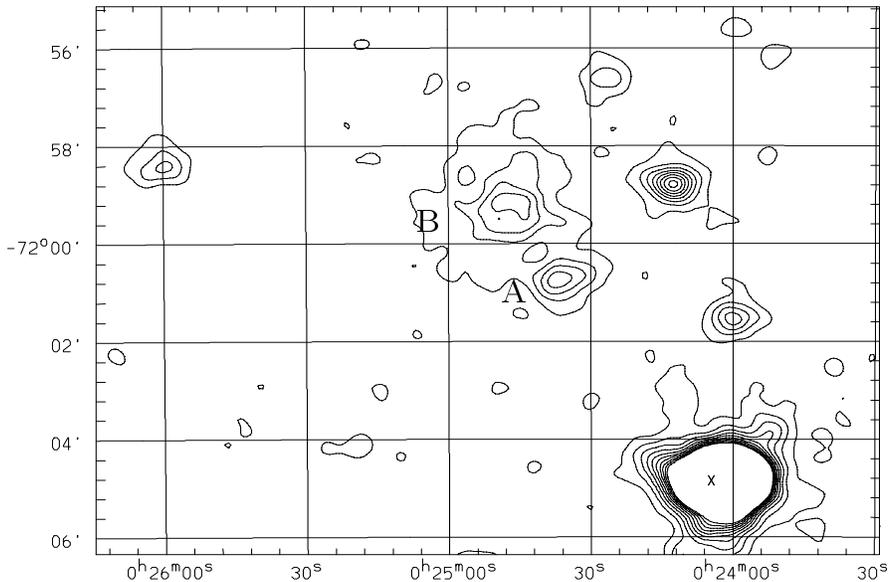

**Figure 3:** Raw contour map of the NE region of the cluster. This image only shows the hard band (including point sources) and has been smoothed with a Gaussian of $\sigma = 9''$. The contour levels start at $1.1 \times 10^{-2} cnts/arcsec^2$, which is 3 times the background. The optical center of the cluster, which is slightly offset from the x-ray center is marked by an "X". There is a striking alignment of the two peaks of the diffuse source, labeled A and B, with the center of the cluster

Figure 3, which is a hard band image, shows two fairly bright diffuse sources that are radially aligned with the cluster center. We will refer to the one that is closest to the center as source A and the outer, larger diffuse source as source B. These two diffuse sources are bright and compact enough to be visible in the raw data. Table 1 shows that both sources A and B are detected significantly in the hard band.

We extracted a spectrum for sources A and B by defining a source region and extracting the pulse height distribution of the events in that region. Using the detector response maps

**Table 1:** Detected diffuse sources

|                | $arcmin^2$ | tot cnts     | source | bg   | $N_\sigma$ |
|----------------|------------|--------------|--------|------|------------|
| Source A soft  |            | $62 \pm 7.9$ | 18     | 44   | 2.3        |
| Source A hard  | 1.3        | $67 \pm 8.2$ | 49     | 18   | 6.0        |
| Source A total |            | $129 \pm 11.4$ | 67   | 62   | 5.9        |
| Source B soft  |            | $248 \pm 15.7$ | 54   | 194  | 3.4        |
| Source B hard  | 5.0        | $285 \pm 16.9$ | 218  | 67   | 12.9       |
| Source B total |            | $533 \pm 23.1$ | 272  | 261  | 11.8       |
| soft diffuse   | 37.3       | $1642 \pm 40.5$ | 177 | 1465 | 4.4        |

**Notes:** Shown are the size of the detection box, total number of counts, source counts, background and significance above background. Derivations of the parameters for each are discussed in the text.

**Table 2:** Results of the spectral fit to a thermal bremsstrahlung spectrum for sources A and B.

|                              | **Source A** | **Source B** |
|------------------------------|--------------|--------------|
| Luminosity [$10^{32}$ erg/sec] | 1.84       | 2.77         |
| Volume [$10^{55}$ cm$^3$]    | 9            | 50           |
| Temperature [$10^6$ K]       | $4.6 \pm 1.5$ | $34 \pm 20$ |
| $n_e$ [$cm^{-3}$]            | 0.75         | 0.24         |
| Mass [$M_\odot$]             | 0.06         | 0.10         |
| Cooling time [$10^7$ yr]     | 1.1          | 9.6          |

**Notes:** Luminosities are calculated for bremsstrahlung spectra with temperatures (with $\pm 2\sigma$ errors) indicated even though sources may be non–thermal. Source volumes assume spherical geometry and source diameters of $1.3'$ and $2.5'$ respectively.

within the IRAF package PROS we converted the pulse height distributions into spectra. We fitted the spectra for sources A and B with three spectral models: bremsstrahlung, Raymond–Smith and powerlaw. Because of the small number of counts we can not distinguish between the models. Raymond–Smith and bremsstrahlung fits give similar temperatures (Table 2). The powerlaw fit gives energy indices of $\sim 1$. To show the difference between sources A and B we show the results of a bremsstrahlung fit in Table 2. The bremsstrahlung fits indicate a temperature of $\sim 3 \times 10^7 K$ for source B and $\sim 5 \times 10^6 K$ for source A. To estimate the errors of these numbers we fited models with varying normalization and temperature to the data. From these fits we calculated a $\chi^2$ grid, from which we derived the errors on the temperature shown in Table 2. These are the 2 $\sigma$ contour values and thus are (approximate) 90 % confidence level uncertainties. Thus the large temperature difference between sources A and B is real.

The bremsstrahlung fit gives fluxes of the sources, which we converted to luminosities (shown in Table 2) assuming both sources are at the distance of 47 Tuc (4.6 kpc). Using the estimated temperature and the bremsstrahlung formula

$$n_e = \sqrt{\frac{L}{1.4 \times 10^{-27} T^{0.5} V}} \qquad (2.1)$$

we made a crude estimate of density, mass and cooling time of the two sources. The volume of each of the sources is estimated by the volume of a sphere with the radius given by the size of each of the sources (diameters of $1.3'$ and $2.5'$ respectively). The results are shown in Table 2.

Below we will explore some possible explanations for this bright diffuse emission. Although we shall conclude the emission may be non–thermal (inverse Compton), the thermal fits of Table 2 allow sources A and B to be compared with each other and the thermal parameters derived below for the bow shock. They also constrain possible thermal models.

### 2.2.2 Possible Cluster of Galaxies

There is a bright diffuse source $18'$ south of the cluster (cf. Figure 1 ). The offset of this diffuse source from the cluster center of $\sim 50$ core radii makes it very unlikely that this source is related to 47 Tuc. In fact, there is reason to believe that it is a extragalactic source: Within an 10 arcmin circle around this source there are nine IRAS point sources found by Gillett et al. (1988). Four of these sources are possibly galaxies which suggests that we are seeing a background cluster of galaxies. We therefore do not include this source in our study of diffuse x–ray sources in 47 Tuc. Below we also show that the diffuse sources A and B, as well as the softer source associated with the bow shock, are unlikely to be additional background galaxy clusters.

### 2.2.3 Possible Detection of Bow Shock

We smoothed the soft band image with point sources removed using a wide Gaussian ($\sigma = 80''$). Figure 2 shows the final contour map of the cluster in the soft band. The highest contour is only about 30 % of the removed constant background in the soft image. Even though the emission is faint, it is suggestive of a flow pattern of hot ($\sim 10^6 K$) gas starting in the NE of the cluster. In this picture there is a bow shock in the NE from which heated gas is flowing in a cone or cylinder around the cluster. Of this flow only two tails are visible due to projection (edge brightening) effects.

Because of the low flux level of this diffuse soft band feature we must consider the possibility of it being due to unresolved point sources, or the harder diffuse sources A and B, or background fluctuations.

It seems unlikely that the extended soft emission is due to a collection of faint, unresolved point sources for two reasons:
i) *The shape of the diffuse emission:* The emission peaks at about 20 core radii, where very few cluster x–ray point sources like cataclysmic variables or millisecond pulsars are expected to be (Grindlay 1994). The emission pattern we would expect to see from a large population of unresolved point sources in the cluster is symmetric around the center and centrally peaked. If the unresolved point sources are not cluster members, i.e. foreground or background sources, then it is very unlikely that they arrange themselves in such a way that they appear like a bow shock with symmetrical flow.

ii) *The spectrum of the diffuse emission:* All of the removed point sources in the image are detected in the hard band as well as in the soft band. Most of the sources are much more prominent in the hard band than in the soft band. Thus the non–detection of any similar extended structure in the hard band makes it very unlikely that the emission is due to point sources.

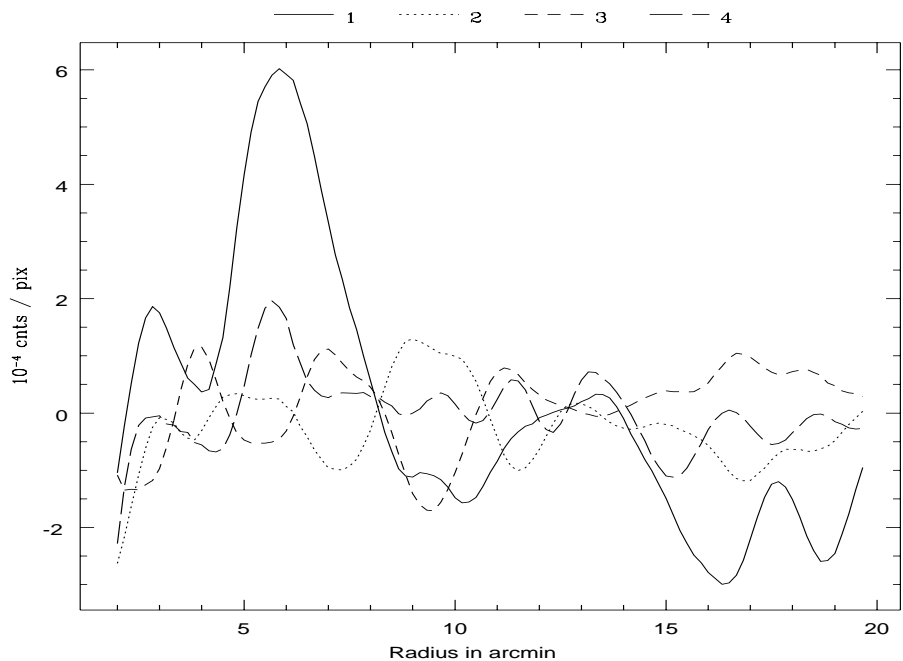

**Figure 4:** Annuli counts in each of the four quadrants of the soft image. The first quadrant is the NE quadrant. The others follow clockwise

Sources A and B, which roughly coincide with the brightest spot in the diffuse soft emission, contribute to the soft diffuse emission shown in Figure 2. In order to test for additional soft diffuse emission we defined the regions of sources A and B in an unsmoothed hard band image by moving a $10'' \times 10''$ box radially outward from the center of each source until the counts in the box were consistent with the background in the hard band. By sliding the box out into different directions we mapped out the spatial extent of sources A and B in the hard band. (The results of this procedure give the source sizes and thus volumes quoted in Table 2). The next step was to exclude the so–defined regions of A and B in the soft band image. In this image we then added counts in concentric annuli around the center of 47 Tuc. The results of these annuli counts, which we broke up into four quadrants, are shown in Figure 4. There is a significant number of soft x–ray photons to the NE of the cluster that are *not* from sources A or B. The numbers for the entry "soft diffuse" in Table 1 were determined from the counts in the annuli between $4'$ and $8'$ in the NE quadrant.

The flatness of the background can be examined qualitatively with Figure 4. There is no strong radial or azimuthal gradient in the number of counts besides the soft diffuse emission in the NE. This suggests that the background is flat all the way in to the cluster center. Neither of the two extended tails produces a signal in Figure 4, which is not surprising because both of them are faint and spread out over several annuli and quadrants.

# 3 Bow shock

The previously reported bow shock in the SE of the cluster by Hartwick et al. (1982) is resolved into six point sources in our much deeper ROSAT image. The temperature and luminosity of this emission were too high to be consistent with a bow shock as pointed out by Faulkner and Smith (1991). The soft diffuse emission we find in our much deeper ROSAT image is a new and better candidate for a bow shock in 47 Tuc. We therefore check the consistency of the soft diffuse emission being due to a bow shock between gas in 47 Tuc and the Galactic halo gas.

## 3.1 Location and Direction

**Table 3:** Proper Motion of 47 Tuc

|  | absolute | | space motion, rotation | | | | space motion, no rotation | | | | projected | |
|---|---|---|---|---|---|---|---|---|---|---|---|---|
|  | $\mu_{RA}$ | $\mu_{dec}$ | $V_{rx}$ | $V_{ry}$ | $V_{rz}$ | $V_{rt}$ | $V_{nrx}$ | $V_{nry}$ | $V_{nrz}$ | $V_{nrt}$ | $\mu'_{RA}$ | $\mu'_{dec}$ |
|  | [mas/yr] | | [km/sec] | | | | | | | | [mas/yr] | |
| Cudworth & Hanson | 3.4 | -1.9 | 31 | -25 | 42 | 58 | -51 | 179 | 42 | 191 | -2.3 | 4.7 |
| Tucholke | 5.5 | -1.6 | -10 | -43 | 32 | 55 | -92 | 161 | 32 | 188 | -0.2 | 5.0 |

**Notes:** The column labeled "absolute" lists the reported measurements. The column "space motion, rotation" is the the LSR space motion of 47 Tuc assuming that the halo gas is co–rotating with the Galaxy (same rotational velocity as the Sun). The first three entries in this column are the three orthogonal velocity components, the fourth is the total velocity. The column "space motion, no rotation" is the same as before, but assuming that the halo gas is not rotating. The last column labeled "projected" is the LSR proper motion vector from the previous entry projected into the equatorial coordinate system.

The first check is the alignment of the direction of the bow shock with the proper motion direction of the cluster. There are two recent measurements of the absolute proper motion of 47 Tuc.

RA: $3.4 \pm 1.7$ mas/yr   dec: $-1.9 \pm 1.5$ mas/yr   (Cudworth and Hanson 1993)
RA: $5.5 \pm 2.0$ mas/yr   dec: $-1.6 \pm 2.0$ mas/yr   (Tucholke 1992)

These numbers are the absolute proper motions of 47 Tuc with respect to the Sun. The relevant proper motion for the location of the bow shock is the motion of the cluster with respect to its own local standard of rest (LSR) defined by the halo gas.

In order to correct for the peculiar motion of the Sun and the Galactic rotation we converted the absolute proper motion of the cluster into Galactic coordinates, which we then transformed into a rectangular coordinate system centered on the Sun, with the x–axis pointing towards the Galactic center, the y–axis pointing in the direction of the Sun's motion around the Galaxy and the z–axis pointing towards the north Galactic pole. In doing so we used the radial velocity of 47 Tuc of –18.8 km/sec (Meylan and Mayor 1986). In the rectangular coordinate frame we corrected for the peculiar motion of the Sun of $(V_x, V_y, V_z) = (9, 12, 7)$ km/sec (Allen, 1973) and the Galactic rotation of the Sun of $(V_x, V_y, V_z) = (0, 220, 0)$ km/sec. The result is the LSR space motion of the cluster, assuming that there is no Galactic rotation at the Galactic height of 3.3 kpc (cf. Table 3, fourth column).

It is not clear what the value of rotation of the halo gas is at a height of 3.3 kpc. To cover the full range of possible values we calculated the LSR space velocity of 47 Tuc under the assumption that the rotation curve of the Galaxy is flat and that the halo gas at the height of 47 Tuc is corotating with the Galactic plane at 220 km/sec. Using a distance of $R_0 = 8.5$ kpc between the sun and the Galactic center we found the LSR space velocity of the cluster (cf. Table 3, third column).

We found that if the halo gas is corotating with the Galaxy, the total LSR velocity of the cluster is very small (∼55 km/sec, cf. Table 3). This velocity is too small to form a bow shock. However, if the halo gas is not rotating the space velocity is ∼200 km/sec (cf. Table 3, fourth column), which is sufficient to produce a bow shock (see discussion in next section). There could be additional velocities in the halo gas in the z direction due to motions connected with the Galactic fountain (Shapiro and Field 1976). However, the data of Danly et al. (1992) indicate that the z-velocity of clouds at 3 kpc above the plane is only about 50 km/sec. Adding such a z velocity to the corotating model does not change the total space motion to the required value of ∼200 km/sec.

Thus the presence of a bow shock would strongly suggest that the rotation in the halo gas at the location of 47 Tuc is very small. We calculated models with different amounts of rotation of the halo gas at the location of 47 Tuc and found that a bow shock can form only if the rotation of the halo gas is ≲ 40 km/sec (0.2 of the Sun's rotation).

Assuming that the halo gas at the location of 47 Tuc is not rotating, we calculated the projected LSR space motions in the equatorial coordinate system for all combinations of absolute proper motions within the quoted errors of the measurements (RA : 1.7 – 7.5 mas/yr, dec : –3.6 – 0.4 mas/yr). The projected RA motions range between –4 and 2 mas/yr. The dec motions we find to be between 3 and 7 mas/yr. The best values that result from the numbers of Tucholke and Cudworth are shown in the fifth column of Table 3. These numbers show that the projected LSR motion of the cluster is mostly towards the north. A possible bow shock would therefore be towards the north of the cluster.

In summary the soft diffuse emission which is seen in the NE of the cluster is consistent with the determinations of the proper motion of the cluster. The estimated space velocity of the cluster with respect to the halo is 200 km/sec, and a slowly or non–rotating halo gas is required.

## 3.2 Temperature and Luminosity

Assuming a halo gas temperature, $T_{in}$, of $10^5 K$ (Savage and de Boer, 1981) and a shock velocity, $v_{in}$, of 200 km/sec we estimated a post–shock temperature of

$$T_{post} = \frac{5v_{in}^2 \bar{m}}{16\gamma k} + \frac{7}{8}T_{in} = 1.4 \times 10^5 \left(\frac{v_{in}}{100 km/sec}\right)^2 + \frac{7}{8}T_{in} \sim 7 \times 10^5 K \qquad (3.1)$$

using the standard strong shock relation where $v_{in}$ is the velocity of the halo gas with respect to the shock and $T_{in}$ the temperature of the halo gas. This temperature is uncertain because the velocity of the cluster and the temperature of the halo gas are not well known. A halo temperature of $\sim 5 \times 10^5 K$ would give a sound speed in the halo that is about the same as the best estimate of the space velocity of 47 Tuc. Thus no shock would form. A detection of a bow shock in 47 Tuc would require that there is at least a component of the halo gas with a temperature of $\sim 10^5 K$ or less.

To get a very rough estimate of the energy in the gas causing the soft diffuse emission we calculated a Raymond–Smith model of a $7 \times 10^5 K$ plasma with solar abundances. Using the standard x–ray spectral package *XSPEC*, we added absorption to the model and folded it with the ROSAT PSPC detector response map. For the absorption we used the 47 Tuc reddening value of E(B − V) = 0.04 (Peterson 1993), which corresponds to $N_H = 4 \times 10^{20}$. The low energy cutoff energy (the energy at which the optical depth $\tau_\nu = 1$) for the assumed hydrogen column is $\sim 170 eV$, which is very close to the temperature of the model. Therefore absorption is very important and the intrinsic flux and luminosity of the source are strongly dependent on the exact values of the temperature of the model and the hydrogen column.

We normalized our Raymond–Smith model to give the observed 177 counts (cf. Table 1) in a 55 ksec ROSAT PSPC exposure. The resulting intrinsic flux was $4.1 \times 10^{-13} erg/sec/cm^2$ in the 0.1 to 2.4 keV band of ROSAT, which corresponds to a luminosity of $8 \times 10^{31} erg/sec$ at the distance of 47 Tuc. This number is very uncertain because it strongly depends on the poorly determined temperature. We therefore only consider it to be an order of magnitude estimate.

We estimated the fraction of the energy that is emitted into the 0.1 − 2.4 keV band by a plasma of $7 \times 10^5$ K to be $\sim 5$ %, using the cooling curves from Raymond et al. (1976). This number again is very sensitive to the assumed temperature. Adopting the value of 5 % we found that the bolometric luminosity of the cooling gas is 20 times higher, or $1.6 \times 10^{33}$ erg/sec (Note that more sensitive soft x–ray observations and emission line measurements are needed to test these predictions. This will be a prime AXAF or XMM investigation). This number must be compared to the amount of kinetic energy available per unit time, given by

$$K = \frac{1}{2}\left(\pi r^2 \varrho_0 v_{in}\right) v_{in}^2 \qquad (3.2)$$

In this equation $r$ is the projected radius of the bow shock and $\varrho_0$ the density of the halo gas. This is just the kinetic energy due to the motion of the cluster through the halo. The actual amount of available kinetic energy is larger, because there is additional kinetic energy in the gas lost from the giants in the cluster. As the velocity at which gas is lost from giants is not well known, we did not include it in the energy balance to get a lower limit on the available kinetic energy. Using likely values we obtained

$$K = 1.9 \times 10^{34} \left(\frac{r}{10pc}\right)^2 \frac{\varrho_0}{1.67 \times 10^{-27} g/cm^3} \left(\frac{v_{in}}{200 km/sec}\right)^3 erg/sec \qquad (3.3)$$

Here $\varrho_0$ is scaled to the assumed halo density $n \sim 10^{-3} cm^{-3}$. Thus the total available kinetic energy is larger by at least an order of magnitude than the total energy lost in radiation.

In their treatment of this problem Faulkner and Smith (1991) estimated that only about $5 \times 10^{-3}$ of the available kinetic energy is radiated by the heated gas. Their assumption was that the halo gas is at rest in the cluster after being shocked. That implies that the acceleration of the halo gas to the velocity of the cluster takes most of the available kinetic energy. However, our results indicate that the halo gas is not at rest after being shocked but instead flowing around the cluster (cf. the "tails" in Figure 3). In this scenario more energy is going into heating the halo gas, which increases the fraction of radiated energy. If the velocity of the shocked halo gas was $v_{in}/4$, as given by the standard strong shock equations, then the amount of kinetic energy in the post shock gas would decrease by a factor of 16. The fraction of radiated energy increases to 8 % of the available energy, which roughly agrees with our estimate of the luminosity of the bow shock and the amount of energy available.

The above considerations show that the diffuse soft emission we find is consistent with the direction of proper motion of 47 Tuc as derived from two recent measurements. We require that the Galactic rotation at the position of 47 Tuc is $\lesssim 0.2$ of the Sun's rotation and that there is a component of the halo gas with T $\lesssim 10^5 K$. With the velocity of 200 km/sec, which results from the proper motion measurements, we constructed a simple bow shock model. The temperature and luminosity we derived from this model are consistent with our ROSAT observation.

## 4 Models for Diffuse Sources A and B

Source B is the most prominent diffuse source in the NE (cf. Figure 3). Of all the diffuse sources in the NE, it has the highest flux and temperature. Indication for cluster membership comes from the rather precise radial alignment of sources A and B with the center of the cluster. It is also peculiar that sources A and B are embedded in the brightest part of a much more extended region of soft diffuse x–ray emission. If the identification of the extended soft x–ray emission with a bow shock is correct, then sources A and B are likely to be connected to the bow shock. In this section we briefly discuss two possible models for sources A and B: a thermal model with an ejected young white dwarf and a non–thermal model due to particle acceleration in the bow shock. We also consider a background galaxy cluster and show this is very unlikely.

### 4.1 Ejected PAGB Star

If sources A and B are cluster members we need either a non–thermal source or a very high velocity wind to heat the surrounding gas to temperatures around $10^7 K$. One possibility is a post Asymptotic Giant Branch (PAGB) star which one expects as end products of stellar evolution in the cluster. Fast winds (∼2000 km/sec) have been observed from central stars of planetary nebulae (PNe) (Patriarchi and Perinotto 1991).

The alignment of sources A and B with the center of the cluster (Figure 3) suggests that an object on a radial orbit might be causing the x–ray emission. The recent discovery of 10 stars with high radial velocities ($\gtrsim$ 30 km/sec) (Pryor et al. 1993) proves the existence of a population of stars that reach the outer regions of the cluster on radial orbits. Thus we consider a model in which the progenitor of a PAGB star leaves the core at 30 km/sec on a radial orbit.

We briefly describe a scenario in which an ejected giant causes the observed x–ray emission in sources A and B: On its way from the core to source B, the giant ejects its envelope. Low mass stars such as the $\simeq 0.8$ $M_\odot$ stars in globular clusters, may not always form a line emitting PN shell, because their central stars evolve too slowly into a hot PAGB star (Kwok 1994). Thus the ejected envelope is dispersed before it can be photoionized by the central star. (We note and suggest that the "optical" PNe known in only two globular clusters, M15 and M22, may instead have formed from more massive mergers such as blue stragglers. A merger could also explain the peculiar absence of H and He emission lines in the spectrum of the PN in M22 (Borkowski et al. 1993)). The motion of the star through the inter–cluster medium decreases the time scale for dispersing the gas and therefore makes it even less likely that a PN shell is formed. When the temperature of the PAGB star gets high enough to drive a fast wind, a bubble of hot ($\sim 10^7 K$) gas is created. Upon crossing the bow shock or another region with increased density this hot bubble is stripped off the central star. The hot gas expands and cools somewhat (yielding the somewhat lower temperature of source A) and is finally confined by the combined gas and magnetic field pressure in the bow shock to form source A. The PAGB star, stripped of all surrounding gas, continues on into the halo. The fast wind ($\sim$ 2000 km/sec) then drives a shock into the halo gas, heating it to $\sim 10^7 K$ to form source B.

The biggest problem with this model is the confinement of the $10^7 K$ hot gas in source B. The timescale for cooling of the gas, if it expands adiabatically, is only $\sim 1000$ yr. The amount of mass accumulated in 1000 yr (wind from the PAGB star plus swept up halo gas) is only around $5 \times 10^{-6} M_\odot$, which is much less than the mass derived from the bremsstrahlung model (Table 2). The only way out is some kind of confinement of the hot gas by either gas or magnetic pressure. Then mass can accumulate over a longer period of time. We could not find any satisfactory way of confining the gas, unless (somehow) swept–up halo magnetic fields can confine the hot gas in a turbulent (tangled) configuration in the bow shock.

In order to make predictions about the current location of the PAGB star we need to estimate its orbital timescale. Assuming the star has slowed down from the 30 km/sec close to the core to 10 km/sec, we find that the star moves 1 pc in $\sim 10^5$ yr. It takes $\sim 5 \times 10^5$ yr at that velocity to move from source A to source B and back. That amount of time is comparable to the duration of the fast wind (Blöcker 1993). Thus the size of sources A and B can be explained, however it is not clear why there is a dark band separating the two sources.

This model was originally partially inspired by a candidate we found (on an old CTIO 4m plate obtained by JEG) for such a PAGB star within the region of source B. However, a spectrum taken with the CTIO 4m identified this candidate to be a foreground F star. Nevertheless, the lack of an optical counterpart in the region of x–ray emission does not rule out this model because if there is a way to confine the hot gas for $\sim 10^6 yr$, then the PAGB star could have moved along its orbit to almost any position in the cluster.

### 4.2 Particle Acceleration in the Bow Shock

The location of sources A and B straddling the bow shock and their alignment with the proper motion of the cluster suggests a physical connection with the bow shock. Inspired by supernova remnant (SNR) shocks, we propose a model in which mildly relativistic electrons, accelerated in the bow shock, inverse Compton (IC) scatter the light of the cluster stars up to x–ray energies. Only mildly relativistic electrons of $\gamma \sim 30$ are needed to raise the energy of optical cluster photons (typically $\sim 2eV$) up to 2 keV. The seed electrons could be provided by the large reservoir of millisecond pulsars (MSPs) in the cluster (Manchester et al. 1991). In the shock acceleration model by Blandford and Eichler (1987) electrons are crossing the shock many times, thereby gaining energy in a first order Fermi process. They are reflected on both sides of the shock by scattering off Alfven waves. These waves can be caused by the relativistic particles themselves, thus requiring no particular magnetic field structure around the shock.

The first order Fermi acceleration of electrons in the bow shock produces a power–law distribution of relativistic electrons. The spectrum of source B can be fit with a power law of (energy) index 1. Assuming an isotropic distribution of photons and electrons, a power law distribution of electrons of the form

$$N_e(\gamma) = C\gamma^{-3} \qquad (4.1)$$

would produce the observed spectrum. The constant factor $C$ can be determined from the measured luminosity using the relation

$$L = \int_{\gamma_{min}}^{\gamma_{max}} P_{IC} N_e(\gamma) d\gamma \qquad (4.2)$$

$P_{IC}$ is the power emitted by a single electron, given by

$$P_{IC} = \frac{4}{3}\sigma_T c \gamma^2 u_{ph} \qquad (4.3)$$

(cf. Rybicki and Lightman 1979), where $u_{ph}$ is the energy density in optical photons. Solving for $C$ we get

$$C = \frac{3L}{4\sigma_T c u_{ph} \ln \frac{\gamma_{max}}{\gamma_{min}}} \qquad (4.4)$$

To determine the energy density in optical cluster photons at the location of Source B we go to a spherical coordinate system with origin at the center of the cluster. Then $u_{ph}$ at position $\vec{r}$ is given by

$$u_{ph}(\vec{r}) = \int \frac{\ell(r')}{4\pi c(\vec{r} - \vec{r'})^2} d^3 r' \qquad (4.5)$$

where $\ell$ is the luminosity density as a function of position. After introducing dimensionless quantities and integration over angles we find

$$u_{ph}(\hat{r}) = \frac{u_c}{\hat{r}} \int_0^{\hat{r}_{max}} \hat{\ell}(\tilde{r}) \ln \left| \frac{\hat{r} + \tilde{r}}{\hat{r} - \tilde{r}} \right| \tilde{r} d\tilde{r} \qquad (4.6)$$

The central photon energy density $u_c$ is given by $u_c = \ell_c r_c / 2c$ and $\hat{r} = r'/r_c$, $\tilde{r} = r'/r_c$, $\hat{\ell} = \ell/\ell_c$. $r_c$ is the core radius and $\ell_c$ is the central luminosity density. The values we adopt for $r_c$ and $\ell_c$ are 0.5 pc and $6.8 \times 10^4 L_\odot/pc^3$ (Djorgovski 1993) respectively. Even though new HST observations (DeMarchi et al. 1993) indicate that 47 Tuc is a post–core collapse cluster, we use a King model (King 1966), which fits the outer parts of the cluster well, to numerically integrate the remaining integral. Evaluating the integral at the position of Source B ($20 r_c$) we get

$$u_{ph}(20 r_c) = 0.025 u_c = 5.6 \times 10^{-12} erg/cm^3 = 3.5 eV/cm^3 \qquad (4.7)$$

With this value and $\gamma_{max} = 30$, $\gamma_{min} = 10$ we find $C = 1.8 \times 10^{57}$. The total energy in relativistic electrons is given by

$$E_{el} = \int_{\gamma_{min}}^{\gamma_{max}} \gamma m_e c^2 N_e(\gamma) d\gamma = 9.8 \times 10^{49} erg \qquad (4.8)$$

The total energy in relativistic electrons is drawn from the kinetic energy available in the bow shock. Assuming that the cluster is swept free of gas during each Galactic plane crossing, the relativistic electrons must be reproduced after each crossing. The time between plane passages is estimated to be a few times $10^8 yr$ from the orbit of 47 Tuc (Tucholke 1992). The current position of 47 Tuc is below the plane (negative galactic latitude). The z–component of its velocity is positive (cf. Table 3), therefore 47 Tuc is moving back towards the plane right now and it has been $\sim 10^8 yr$ since the cluster last crossed the plane.

Assuming that as much as 50 % of the available kinetic energy (eq. (3.3)) is going into particle acceleration in the bow shock, we find that it takes $3.4 \times 10^8 yr$ to get to the total energy of eq. (4.8). Thus there might have been just enough time to build up the required energy in relativistic electrons. Also note that eq. (3.3) is a conservative estimate of the available energy.

A second consistency check is the required total number of relativistic electrons, given by

$$N_{tot} = \int_{\gamma_{min}}^{\gamma_{max}} N_e(\gamma) d\gamma = 9 \times 10^{54} \qquad (4.9)$$

As the temperature of the post–shock gas is only $7 \times 10^5$ K, we cannot assume that there are enough relativistic seed electrons from the hot gas. We therefore assume that the seed electrons are provided by the population of MSPs in the cluster. Using a conservative estimate of the spin down energy loss rate, $\dot{E}$, of a MSP of $\sim 10^{33} erg/s$ (vs. the spin down luminosity of PSR1957+20, which is $\sim 10^{35} erg/s$), and assuming that $\sim$10 % of this energy goes into a relativistic wind of electrons with Lorentz factors $\gamma \sim$ 1–3 (cf. Kulkarni et al. 1991), then the particle luminosity of a MSP is $\sim 10^{32} erg/sec$. Using these numbers we estimate the number of such electrons emitted from a MSP to be $4 \times 10^{37} e^-/sec$. Thus in $3.4 \times 10^8 yr$ a single MSP emits $4 \times 10^{53} e^-$. Assuming that the bow shock covers 10 % of the total surface area of a sphere of 20 $r_c$ radius around the center of the cluster and that all the MSPs are interior to the bow shock, and that the relativistic electrons are able to diffuse out uniformly, a total number of $\sim$ 200 MSPs is required. Clearly all of these numbers are very approximate, and the assumptions of "typical" $\dot{E}$ and free diffusion of these low energy cosmic ray electrons through the cluster are questionable and need to be considered in much greater detail than appropriate for this paper. However, it appears that a plausible total number of MSPs in 47 Tuc (given the 10 MSPs seen by Manchester et al. 1991) could give rise to the relativistic electron flux required.

It is not clear why the spatial extent of sources A and B is much smaller than the bow shock. Possibly particle acceleration is limited to the center of the bow shock where the shock velocity is largest. The formation of two sources of different size on either side of the bow shock could be due to the different magnetic field strengths ahead and behind the bow shock. These questions about the location and shape of sources A and B need more detailed modeling.

### 4.3 Galaxy Cluster

Given the high temperature of source B, a galaxy cluster might seem a likely explanation. On the other hand it seems unlikely to find a galaxy cluster so close to 47 Tuc.

In order to quantify the likelihood of source B being a galaxy cluster we use the results of the Einstein Medium Sensitivity Survey (EMSS). The EMSS produced a $\log N - \log S$ plot for galaxy clusters (Gioia et al. 1984). Recent ROSAT deep survey results suggest that the EMSS galaxy cluster surface density is over–estimated (Bower et al. 1994) because galaxy clusters formed recently. Using the EMSS results as a conservative upper limit we find the chance of finding one or more galaxy clusters in a $10'$ circle around the center of 47 Tuc is 5.8% . This probability is small, but not small enough to exclude the possibility.

In order to search for possible cluster galaxies we have analyzed a blue plate (emulsion IIIaJ, filter GG 385) taken by W. Liller on Jan 21 1979 on the 4m telescope at CTIO. The calibration curve indicates that the limiting magnitude on the CTIO plate is somewhere around 22. A star of 19th mag is a well defined object and is a conservative limit for resolving background objects from cluster stars at the position of source B. There is no obvious galaxy, or diffuse object, in the region of source B. We also looked for galaxies south of the cluster, to check whether we can find the IRAS point sources that are most likely galaxies (cf. discussion above). At the position of the IRAS sources we find three faint, diffuse objects, suggesting that indeed we can identify galaxies, and thus a galaxy cluster, on this plate.

A distant galaxy cluster ($z > 0.2$) cannot be ruled out totally, but it seems highly unlikely that such a galaxy cluster falls within other regions of diffuse x–ray emission likely to be associated with the globular cluster (i.e. the bow shock). The probability of a galaxy cluster aligned with a bow shock in a globular cluster would be much smaller than the probability estimated above: only $\lesssim 0.5$ % given the projected area of the bow shock. There is further the unlikely alignment of sources A and B, which are separated by $2'$ and lying on a straight line through the center if the cluster. Thus we conclude that source A and B are *not* likely to be a galaxy cluster.

The best test to rule out a galaxy cluster as a source of the emission is by obtaining a deep spectrum, in which we could measure the redshift of the 6.7 keV iron lines. The limited spatial resolution of ASCA means this test will likely require AXAF. We have also proposed deep $H\alpha$ and R images for HST (using WFPC2) to study the bow shock. The R images would also enable us to conduct a conclusive test for a galaxy cluster at the position of source B and $H\alpha$ images would enable a study of the bow shock and its possibly associated cooler and denser filament structure.

## 5 Conclusions

We searched a deep ROSAT PSPC image of 47 Tuc for a diffuse x–ray emission. We found soft emission whose location, luminosity and temperature is consistent with a bow shock model. However, the emission is very faint and only visible after careful background and point source removal. The picture that results from a bow shock model is that the projected LSR space motion of 47 Tuc is towards the NE. There is a bow shock at a radius of about $6'$ in the NE between cluster gas and halo gas. The required velocity to form a bow shock of $\sim$200 km/sec constrains the rotation of the halo gas at the position of 47 Tuc to be $\lesssim$ 40 km/sec. The presence of a bow shock also requires a component of the halo gas with a temperature of $\lesssim 10^5 K$. The post shock temperature is about $7 \times 10^5$ K and the total shock x–ray luminosity about $8 \times 10^{31} erg/sec$ (in the ROSAT band). The estimated total shock luminosity is $1.6 \times 10^{33} erg/sec$, which is only $\sim$ 10 % of the available kinetic energy.

Identification of the diffuse soft emission with a bow shock is complicated by the poorly understood sources A and B. Source A is lying within the brightest area of the bow shock and source B just outside of it. The spectra of sources A and B indicate a much harder spectrum than we expect and infer for the bow shock. The soft diffuse emission of the bow shock is only detected in the lowest energy bands (0.1 – 0.4 keV) which agrees with a plausible shock temperature of $7 \times 10^5 K$. Sources A and B are detected significantly only in the hard energy band (0.4 – 2.4 keV). Their estimated temperatures, if due to hot gas, are $4.6 \times 10^6 K$ and $3.4 \times 10^7 K$ respectively. Even though they coincide with the soft diffuse emission, they are distinct in size and temperature.

We proposed a model for all three detected diffuse sources in the NE which identifies the soft diffuse emission with a bow shock, and sources A and B as inverse Compton sources due to electrons accelerated in the bow shock. This model explains in a consistent way the observed luminosities and positions of the three sources. More detailed modeling of the bow shock and particle acceleration is needed to further test this model. A thermal model for sources A and B, which involves a high velocity wind from a PAGB star on a radial orbit, is also possible but leads to hot gas confinement problems. Additional observations with HST and eventually AXAF can map all three sources in detail. This might then enable even better constraints on the rotation and temperature of the galactic halo as well as the total number of MSPs in 47 Tuc. The more sensitive AXAF observations will be of great interest to map the bow shock "tails", and the flow of gas out of the cluster, in detail.

We thank B. Balick and J. Raymond for discussions and C. Bailyn and J. McClintock for obtaining optical data on a PAGB candidate. This work was supported by grants NAG5–1624 and NAGW–3280.